\documentclass[a4paper]{jpconf}
\usepackage{graphicx}
\begin{document}
\title{LHCb Topological Trigger Reoptimization}

\author{Tatiana Likhomanenko$^{1, 2, 3}$, Philip Ilten$^{5}$, Egor Khairullin$^{1, 4}$, Alex Rogozhnikov$^{1, 2}$, Andrey Ustyuzhanin$^{1, 2, 3, 4}$, Michael Williams$^{5}$}

\address{$^1$ Yandex School of Data Analysis (YSDA), RU}
\address{$^2$ National Research University Higher School of Economics (HSE), RU}
\address{$^3$ NRC "Kurchatov Institute", RU}
\address{$^4$ Moscow Institute of Physics and Technology, Moscow, RU}
\address{$^5$ Massachusetts Institute of Technology, US}

%\address{Production Editor, \jpcs, \iopp, Dirac House, Temple Back, Bristol BS1~6BE, UK}

\ead{tatiana.likhomanenko@cern.ch, mwill@mit.edu}

\begin{abstract}
The main b-physics trigger algorithm used by the LHCb experiment is the so-called topological trigger. The topological trigger selects vertices which are a) detached from the primary proton-proton collision and b) compatible with coming from the decay of a b-hadron. In the LHC Run 1, this trigger, which utilized a custom boosted decision tree algorithm, selected a nearly 100\% pure sample of b-hadrons with a typical efficiency of 60-70\%; its output was used in about 60\% of LHCb papers. This talk presents studies carried out to optimize the topological trigger for LHC Run 2. In particular, we have carried out a detailed comparison of various machine learning classifier algorithms, {\em  e.g.}, AdaBoost, MatrixNet and neural networks. The topological trigger algorithm is designed to select all "interesting" decays of b-hadrons, but cannot be trained on every such decay. Studies have therefore been performed to determine how to optimize the performance of the classification algorithm on decays not used in the training. Methods studied include cascading, ensembling and blending techniques. Furthermore, novel boosting techniques have been implemented that will help reduce systematic uncertainties in Run 2 measurements. We demonstrate that the reoptimized topological trigger is expected to significantly improve on the Run 1 performance for a wide range of b-hadron decays.\end{abstract}

\section{Introduction}
The LHCb trigger is divided into three stages: a hardware, or level-0 (L0) stage, and two software, or high-level, stages (HLTs) \cite{trigger_system}. The second stage of the HLT (HLT2) processes few enough events that it is possible to perform reconstruction that is very similar to what is done offline. This allows the HLT2 to use multivariate algorithms. There are many HLT2 lines dedicated to triggering on various types of events. This note provides a detailed description of the HLT2 topological lines reoptimized for LHC Run 2. Most n-body hadronic B decays $(n \geq 3)$ are only triggered on efficiently in LHCb by these lines. This note also presents a new HLT scheme for LHC Run 2, which includes a more widespread usage of multivariate algorithms.  The HLT2 LHC Run 1 is described in details in \cite{topo1}.

\section{HLT LHC Run 2 scheme}
\begin{figure}[h]
\includegraphics[width=16pc]{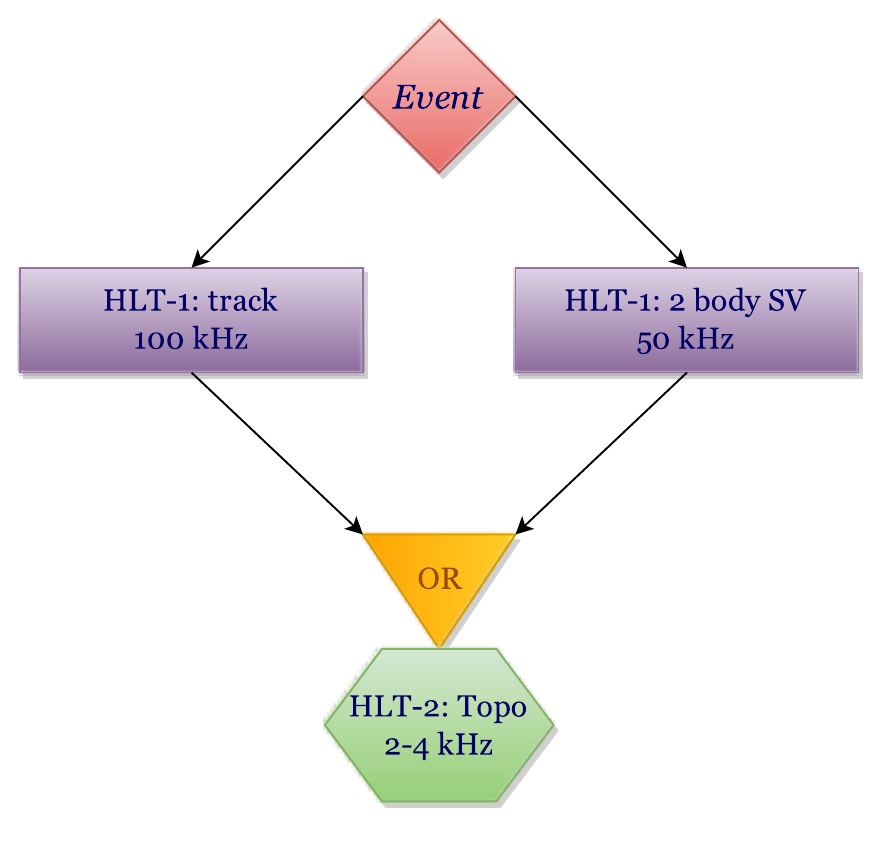}\hspace{2pc}%
\begin{minipage}[b]{20pc}\caption{\label{scheme} HLT Run 2 scheme. HLT1 consists of two lines: the "1-track" and "2 body SV", with corresponding output rates 100 kHz and 50 kHz. If an event passes at least one of these two lines, then it is triggered by the topological line with output rate 2-4 kHz.}
\end{minipage}
\end{figure}

In the HLT1 during Run 1, the so-called  "1 track" line was used because there was not sufficient CPU time for combinatorics and because the tracking transverse momentum threshold was too high for an secondary vertex (SV) based approach to be efficient. In Run 2, the HLT1 tracking transverse momentum threshold will be the same as was used in the HLT2 for Run 1.  This makes it possible to use, along with a "1 track" line, a "2 body SV" line. Thus, we consider a new HLT scheme (figure~\ref{scheme}). The combination of displacement from the primary vertex (PV) and high transverse momentum are the main characteristics of interesting decays. The HLT "1-track" line is looking for either one super high transverse momentum or high displacement track. The HLT "2 body SV" line uses a multivariate analysis to look for two tracks that form a vertex. The HLT topological line is improved using a more powerful multivariate analysis. It uses fully reconstructed events to look for 2, 3, and 4 tracks making a vertex.

The analysis was conducted on the following data: signal samples are simulated 13-TeV B-decays of various topologies, while the background sample is generic Pythia 13-TeV proton-proton collisions.

\section{Multivariate Analysis}
\begin{figure}[h]
\includegraphics[width=22pc]{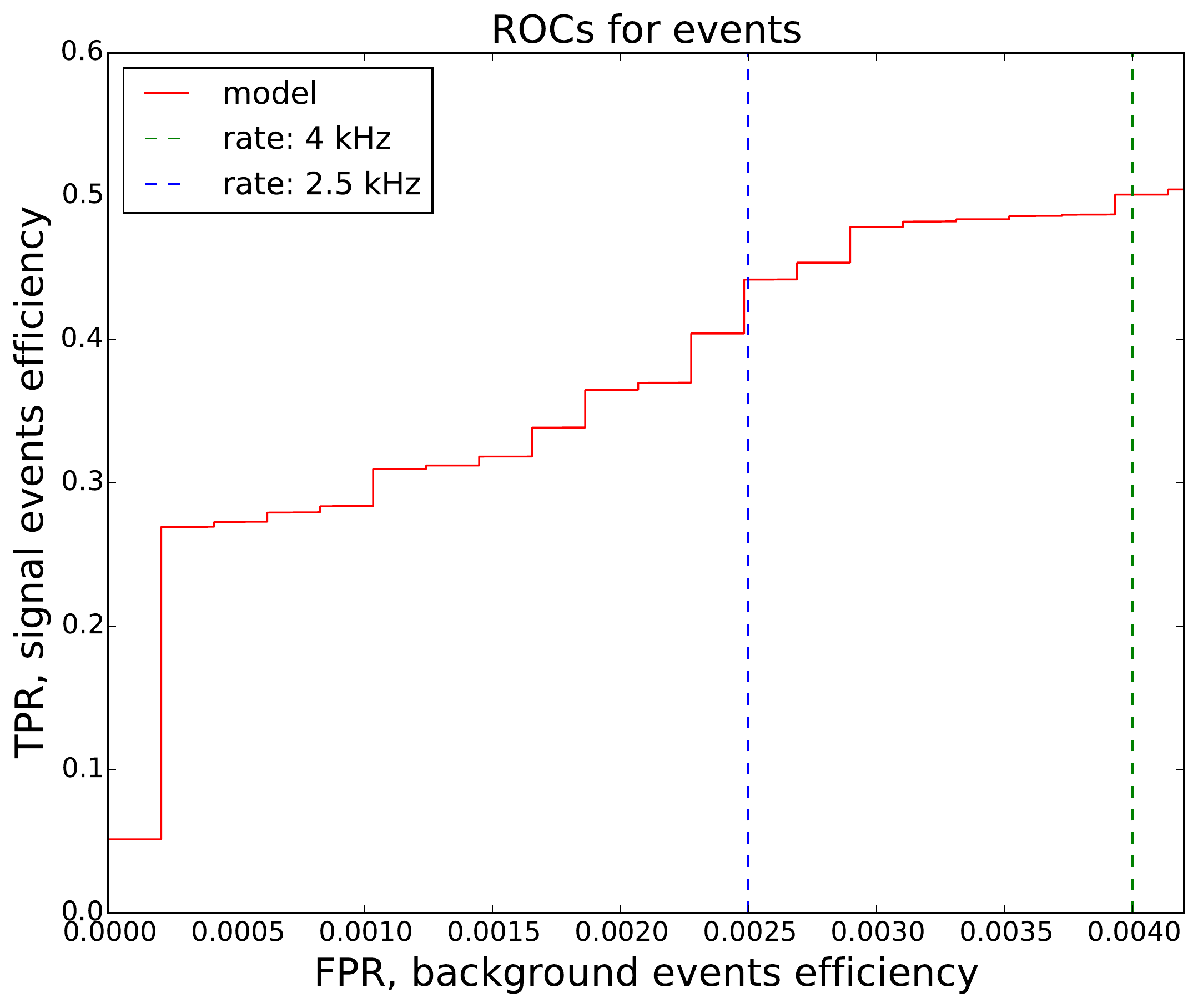}\hspace{2pc}%
\begin{minipage}[b]{14pc}\caption{\label{roc} Trigger events ROC curve. An output rate of 2.5 kHz corresponds to an FPR of 0.25\%, 4 kHz~--- 0.4\%. Thus to find the signal efficiency for a 2.5 kHz output rate, we take 0.25\% background efficiency and find the point on the ROC curve that corresponds to this FPR.}
\end{minipage}
\end{figure}

Each event is represented as set of secondary vertices. These secondary vertices are the input data to the classifier.  The event is triggered if at least one secondary vertex passes the classifier selection. A trigger's output rate is limited. This restriction is equivalent to a restriction on the background events efficiency (FPR). For example, for an output rate of 2 kHz, then FPR is 0.2\%. This can be shown using ROC curves from the classifier, which are plotted for test events (figure~\ref{roc}).

All classifiers were trained on one half of the data and were tested on the other half. For the HLT "1-track" and "2 body SV" lines, all signal modes were used, while for the HLT topological line six specific modes were used in training.  All algorithms were tested on all available signal modes.

\subsection{HLT "1-track" line}
\begin{figure}[h]
\begin{minipage}{18pc}
\includegraphics[width=18pc]{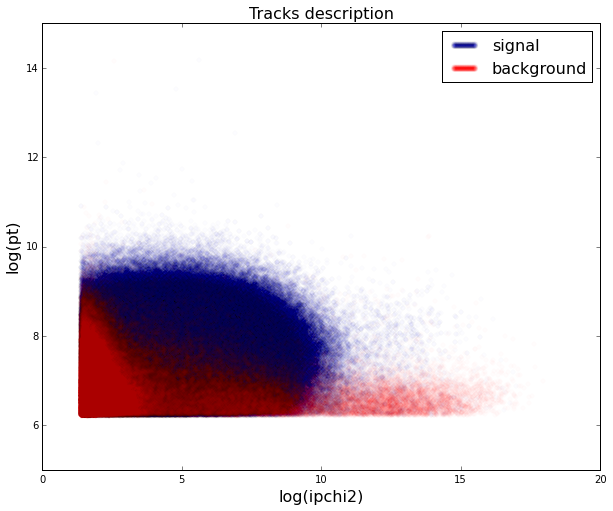}
\caption{\label{hlt1} Track data scatters, described in two-dimensional space.}
\end{minipage}\hspace{2pc}%
\begin{minipage}{18pc}
\includegraphics[width=18pc]{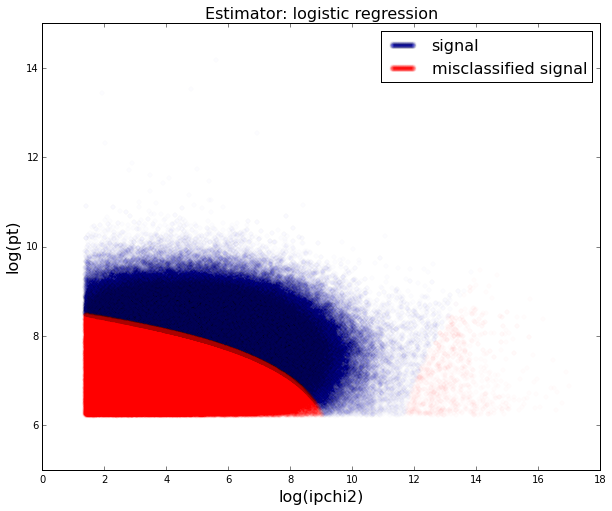}
\caption{\label{hlt1log} Decision boundary for logistic regression.}
\end{minipage} 
\begin{minipage}{18pc}\vspace{1pc}%
\includegraphics[width=18pc]{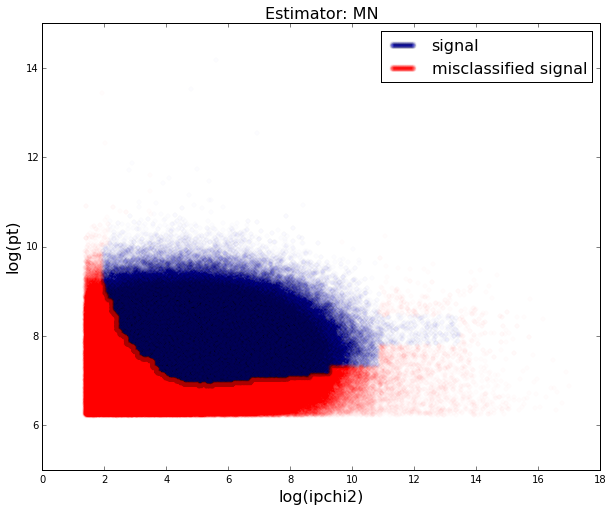}
\caption{\label{hlt1mn} Decision boundary for MatrixNet algorithm.}
\end{minipage}\hspace{2pc}%
\begin{minipage}{18pc}\vspace{1pc}%
\includegraphics[width=18pc]{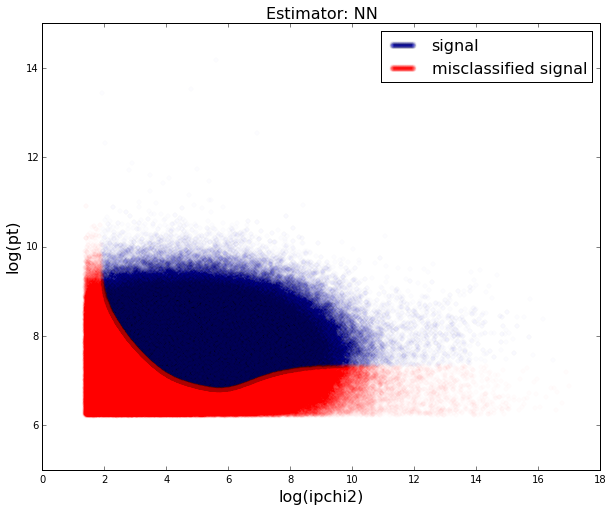}
\caption{\label{hlt1nn} Decision boundary for neural networks.}
\end{minipage} 
\end{figure}

Preselections and variables for the HLT "1-track" line are listed in table~\ref{hlt1sel}. Only two variables are used in this line. To optimize the signal efficiency and find an optimal decision boundary,  multivariate analysis is used. Distributions of signal and background are shown in figure~\ref{hlt1}. The MatrixNet\cite{mn_paper} algorithm (figure~\ref{hlt1mn}), logistic regression (figure~\ref{hlt1log}) and neural networks (figure~\ref{hlt1nn}) were trained to find the decision boundaries. From experiments, MatrixNet decision boundary is found to be the best, but for online processing a more simple decision rule fits: a hyperbolic function can be used as the decision boundary. Efficiencies comparison of different algorithms for B-modes is shown in figure~\ref{hlt1_res}.

\begin{center}
\begin{table}[h]
\centering
    \caption{\label{hlt1sel} HLT "1-track" line description.}
    %\footnotesize\rm
    \centering
    \begin{tabular}{@{}*{7}{l}}
    \br
    Track preselections: & \\
    
    \verb  & $PT > 500$ MeV\\
    \verb  & $IP_{\chi^2} > 4$\\
    \verb  & $track_{\chi^2}/ndof < 3$\\
    \br
    Analysis variables: & $PT$, $IP_{\chi^2}$ \\
    \br
    Output rate: & 100 kHz\\
    \br
    \end{tabular}
\end{table}
\end{center}

\subsection{HLT "2 body SV" line}
\begin{figure}[h]
\begin{minipage}{18pc}
\includegraphics[width=18pc]{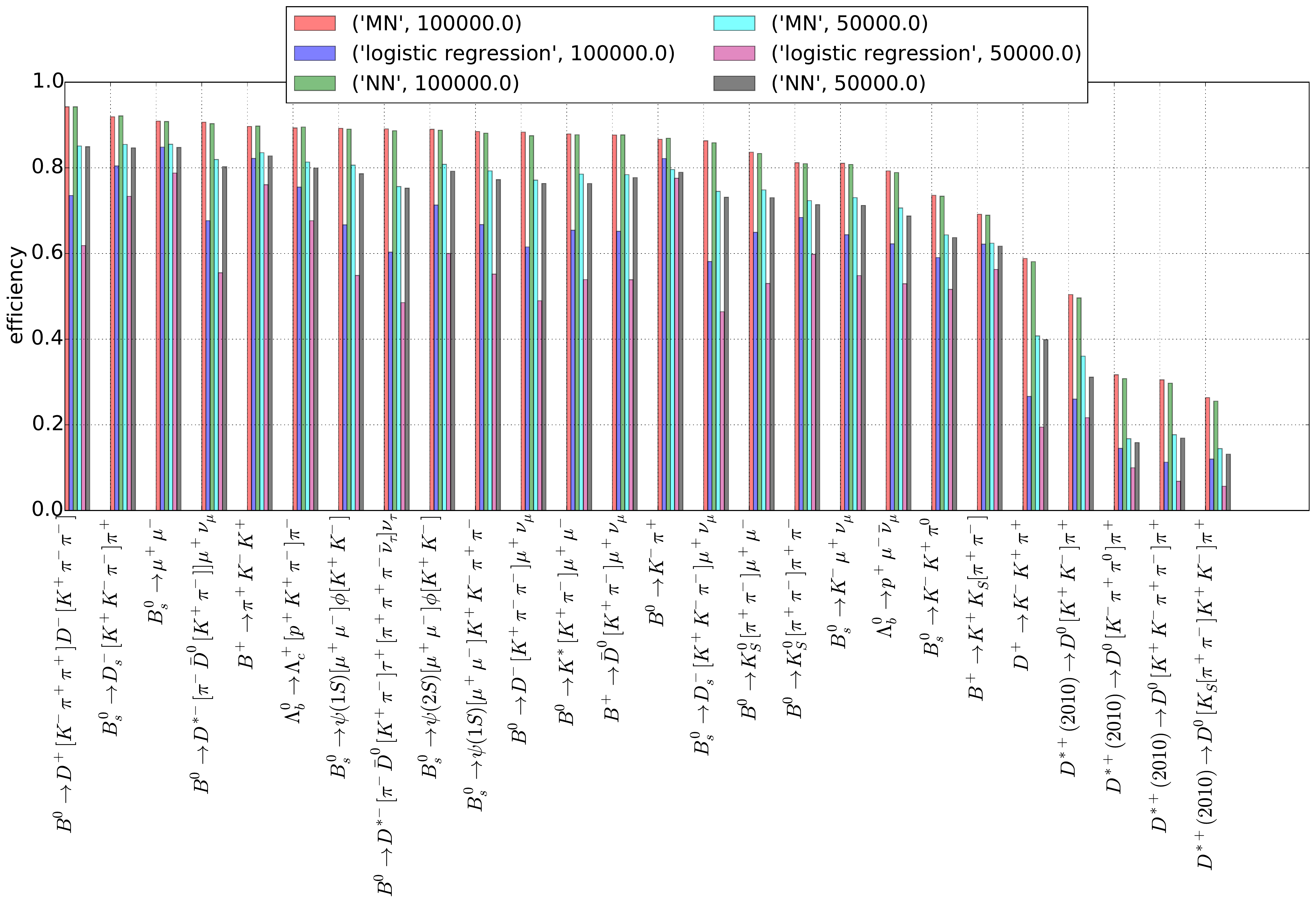}
\caption{\label{hlt1_res} Efficiencies comparison for MatrixNet (MN), neural networks (NN) and logistic regression. 50 kHz and 100 kHz output rates are considered. \\ \\ \\ }
\end{minipage}\hspace{2pc}%
\begin{minipage}{18pc}
\includegraphics[width=18pc]{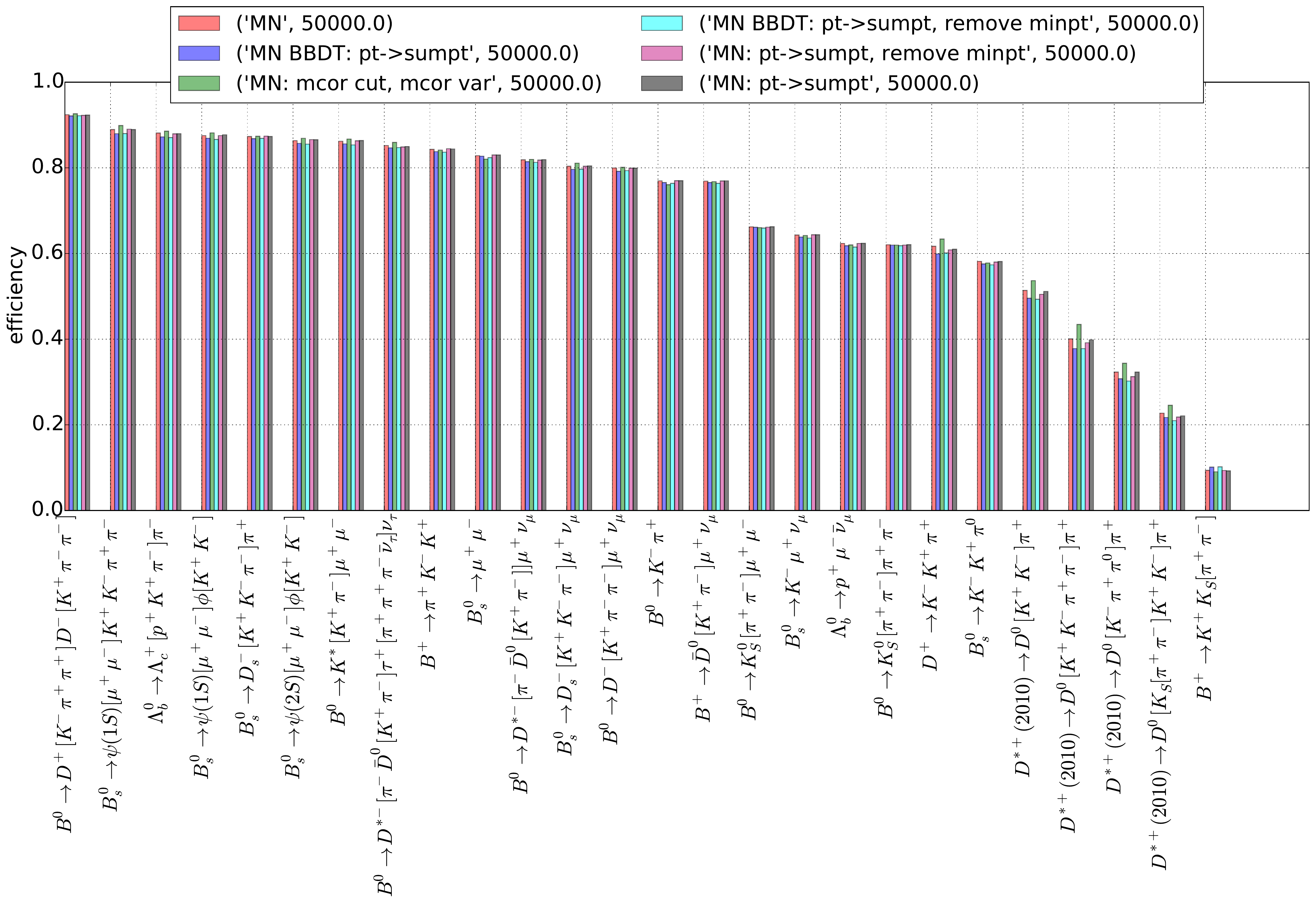}
\caption{\label{hlt12_res} Efficiencies comparison for MatrixNet: with corrected mass, without it (MN), with the scalar sum of transverse momentum (sum $pt$, without min and vector $pt$). These models were `converted' to BBDT format and compared. The output rate is set to 50~kHz.}
\end{minipage} 
\end{figure}

Preselections and variables for the HLT "2 body SV" line are listed in table \ref{hlt1svsel}. The line looks for two tracks that form a vertex. In this case, the MatrixNet algorithm is used and several studies were conducted. Firstly, the possibility to remove the corrected mass cut ($mcor<10$) and to remove the corrected mass as variable from classifier's input was investigated.  The impact on the performance is negligible.  These removals were done to reduce systematic uncertainties and to help in exotic searches. Secondly, to minimize the set of variables used in trigger, a selection was conducted among the scalar sum, vector sum, and minimum of the transverse momenta. The results obtained (efficiencies comparison) are shown in figure \ref{hlt12_res}.

\begin{table}[h]
%\centering
\begin{minipage}{.48\textwidth}
  %\centering
    \caption{\label{hlt1svsel} HLT "2 body SV" line description.}
    %\footnotesize\rm
    \centering
    \begin{tabular}{@{}*{2}{l}}
    \br
    Track preselections: \hspace{-1cm} & \\
    
    \verb  & $PT > 500$ MeV\\
    \verb  & $IP_{\chi^2} > 4$\\
    \verb  & $track_{\chi^2}/ndof < 2.5$\\
    \br
    SV preselections: \hspace{-1cm} & \\
    
    \verb  & $PT > 2$ GeV\\
    \verb  & $vertex_{\chi^2} < 10$ \\
    \verb  & $1 < mcor$ GeV \\
    \verb  & $2 < \eta < 5$ (PV to SV) \\
    \br
    Analysis variables: \hspace{-1cm} & \\
    \verb  & sum $PT$, $vertex_{\chi^2}$, $FD_{\chi_2}$,  \\
    \verb  & $N$(tracks with $IP_{\chi^2} < 16$) \\ 
     \verb  & \\ 
      \verb  & \\ 
    \br
    Output rate: \hspace{-1cm} & 50 kHz\\
    \br
    \end{tabular}
  \end{minipage}
%\centering  
%\hspace{0.5pc}
\begin{minipage}{.45\textwidth}
  \caption{\label{hlt2sel} HLT topological line description.}
%\footnotesize\rm
   % \centering
    \begin{tabular}{@{}*{2}{l}}
    \br
    Track preselections: \hspace{-1cm} & \\
    
    \verb  & $PT > 200$ MeV\\
    \verb  & $IP_{\chi^2} > 4$\\
    \verb  & $track_{\chi^2}/ndof < 2.5$\\
    \br
    SV preselections: \hspace{-1cm} & \\
    
    \verb  & $vertex_{\chi^2} < 10$ \\
    \verb  & $1 < mcor < 10$ GeV \\
    \verb  & $2 < \eta < 5$ (PV to SV) \\
    \verb  & $N$(tracks with $IP_{\chi^2} < 16) < 2$\\
    \br
    Analysis variables: \hspace{-1cm} & \\
    \verb  & n, mcor, sum $PT$, $vertex_{\chi^2}$, \\
    \verb  & $\eta$, $FD_{\chi_2}$, min $PT$, \\
    \verb  & $IP_{\chi^2}$, $N$(tracks), \\
    \verb  & $N$(tracks with $IP_{\chi^2} < 16$) \\ 
    \br
    Output rate: \hspace{-1cm} & 2-4 kHz\\
    \br
    \end{tabular}
  \end{minipage}
\end{table}

\subsection{HLT topological line}
The HLT2 topological lines are designed to trigger efficiently on any B decay with at least 2 charged children. It is designed to handle the possible omission of child particles.  To save CPU time in the HLT2 reconstruction, only tracks with transverse momentum $PT > 200$ MeV are reconstructed. To reduce the background rate due to ghosts, all tracks are required to have a $track_{\chi^2}/ndof$ value less than 2.5. To reduce the background rate due to prompt particles, all tracks are required to have an impact parameter $\chi^2$ value greater than 4. Due to the inclusive nature of the HLT2 topological lines, this does not mean that all of the B daughters need to satisfy these criteria. The trigger is designed to allow for the omission of one or more daughters when forming the trigger candidate. The previous topological line for Run 1 is described in \cite{topo_2}. 

\begin{figure}[h]
\begin{minipage}{18pc}
\includegraphics[width=18pc]{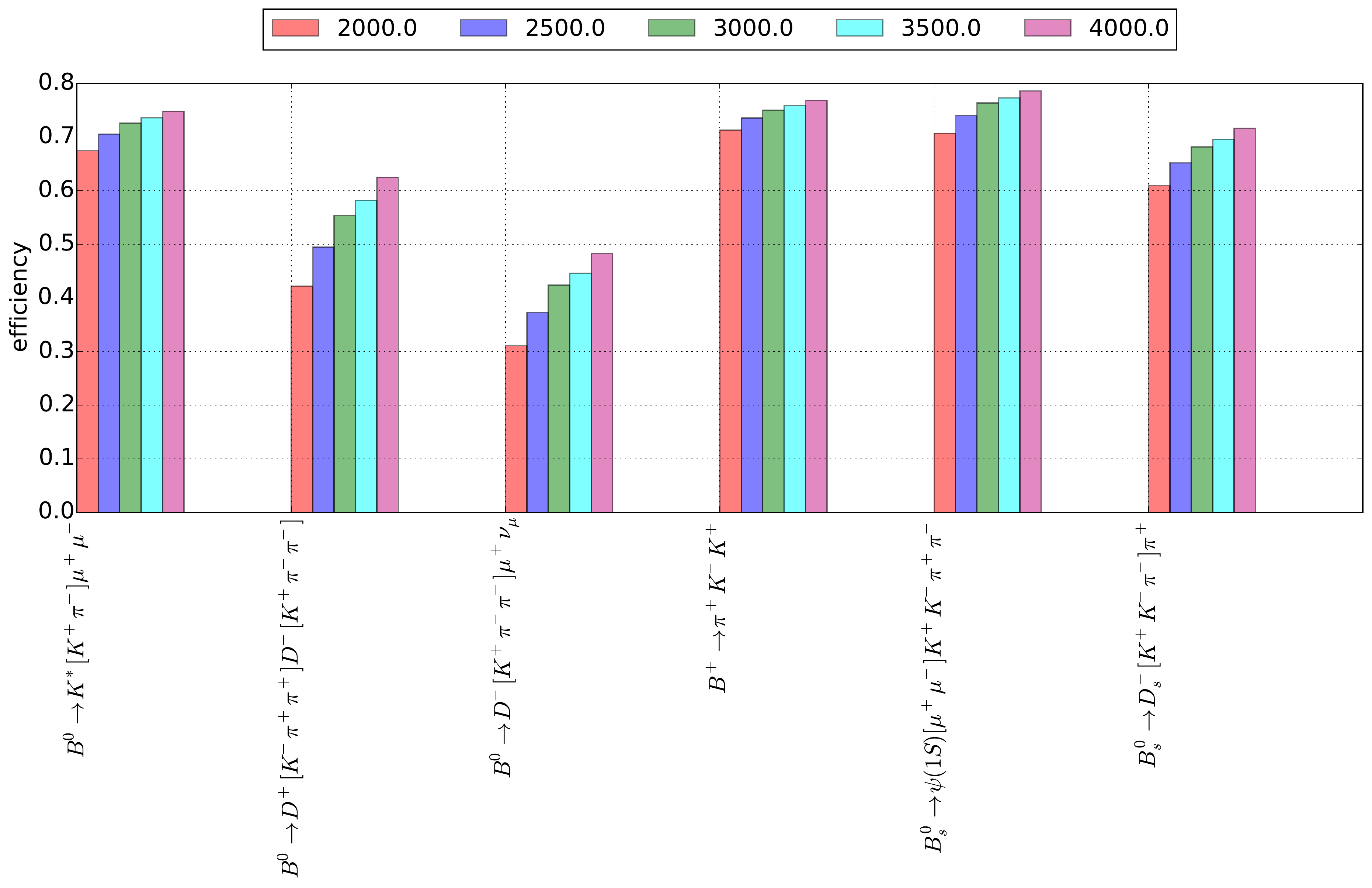}
\end{minipage}\hspace{2pc}%
\begin{minipage}{18pc}
\vspace{+0.4cm}
\includegraphics[width=18pc]{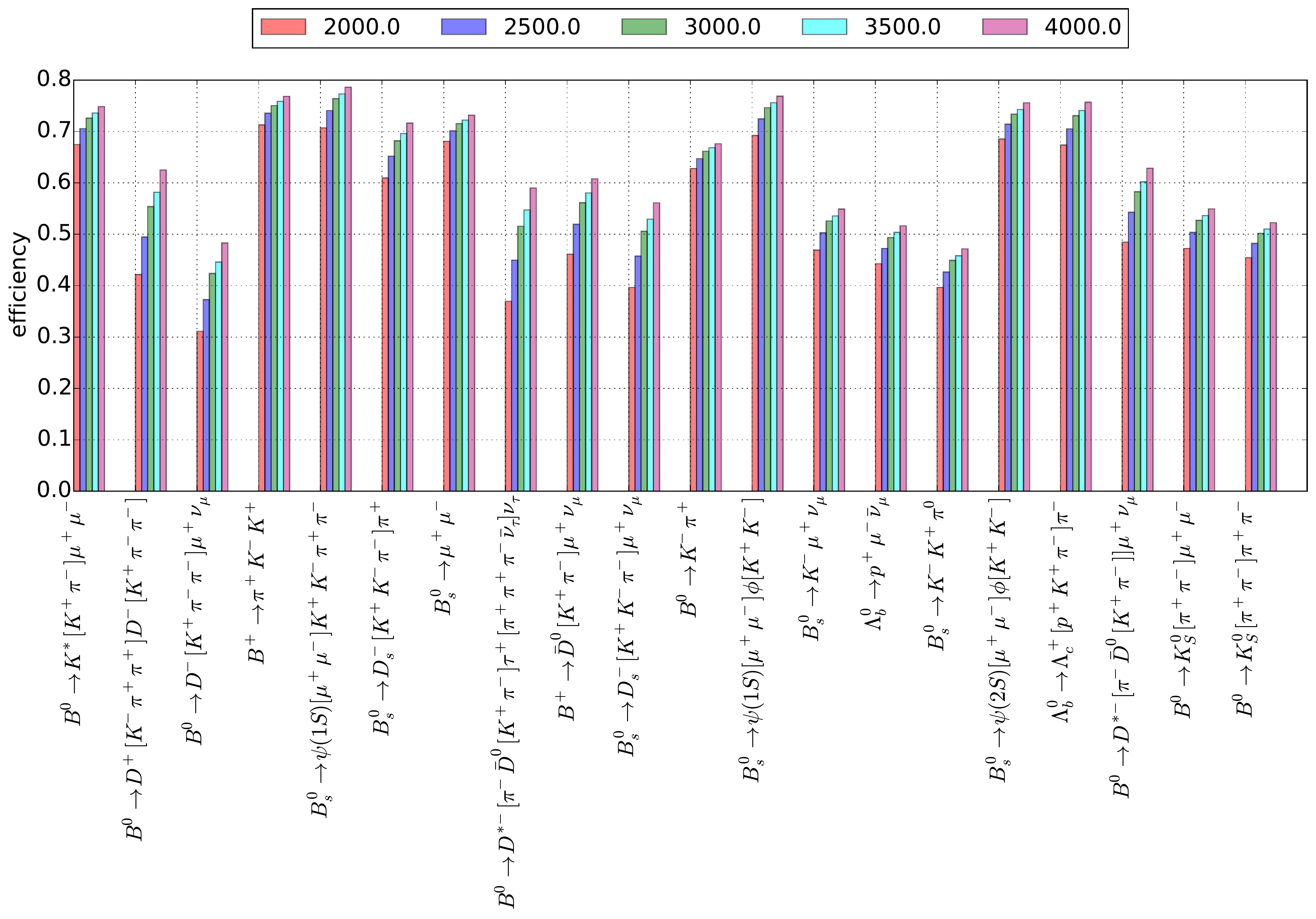}
\end{minipage} 
\caption{\label{hlt2_out} Signal efficiencies for training modes (left) and other available modes (right) for different output rates.}
\end{figure}

In the HLT topological line used in Run 1, a simple boosted decision tree was used \cite{topo_2} to define interesting secondary vertices. For Run 2, the algorithm is reoptimized.  Current preselections and training variables are listed in table \ref{hlt2sel}. The output rate is one of the signal efficiencies factor.  The efficiency dependence on the output rate is shown in figure \ref{hlt2_out}. Thus, several modes, including two training ones, significantly depend on the output rate. 

\begin{figure}[h]
\begin{minipage}{18pc}
\vspace{1cm}
\includegraphics[width=18pc]{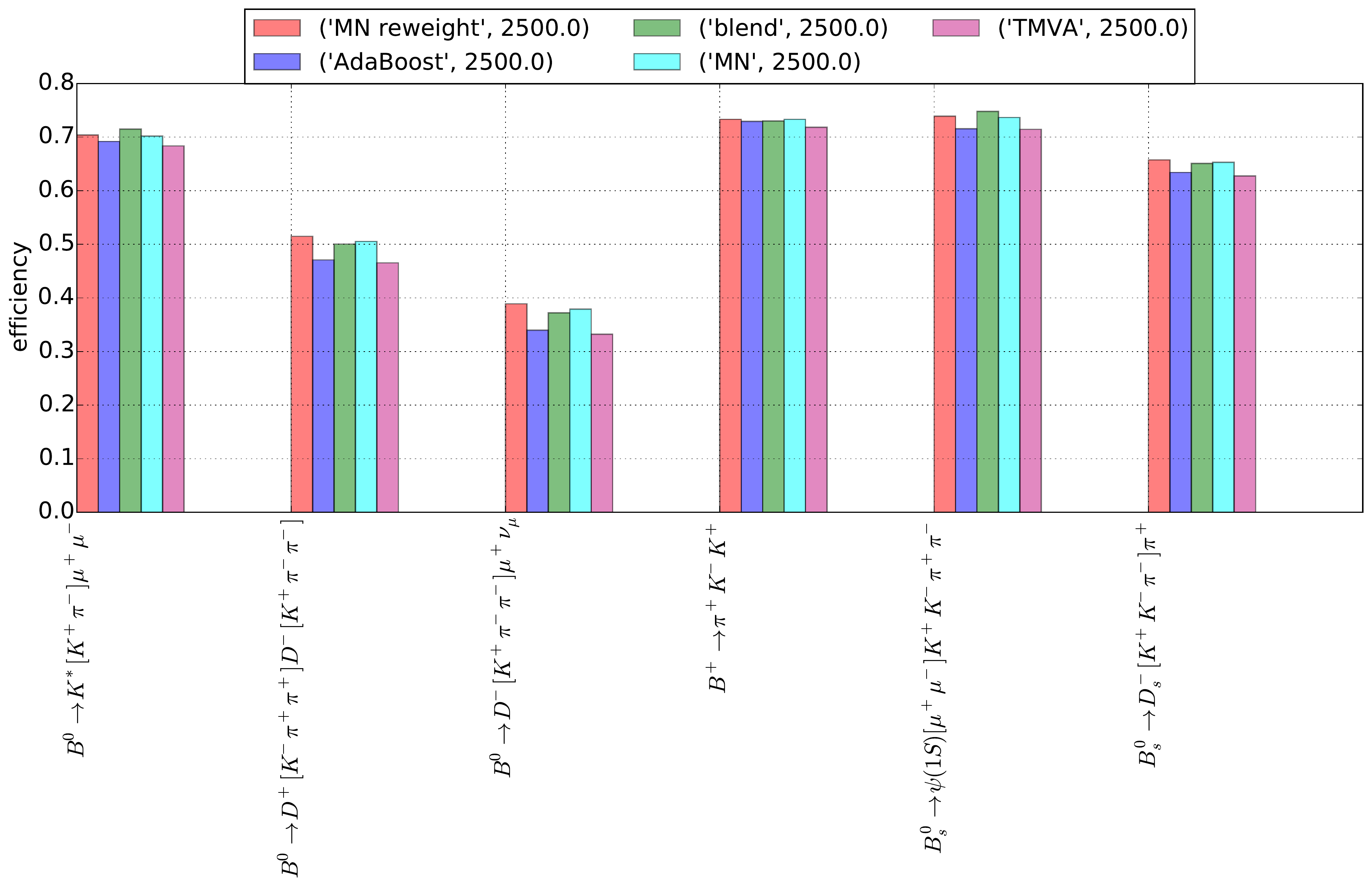}
\end{minipage}\hspace{2pc}%
\begin{minipage}{18pc}
\vspace{-0.3cm}
\includegraphics[width=18pc]{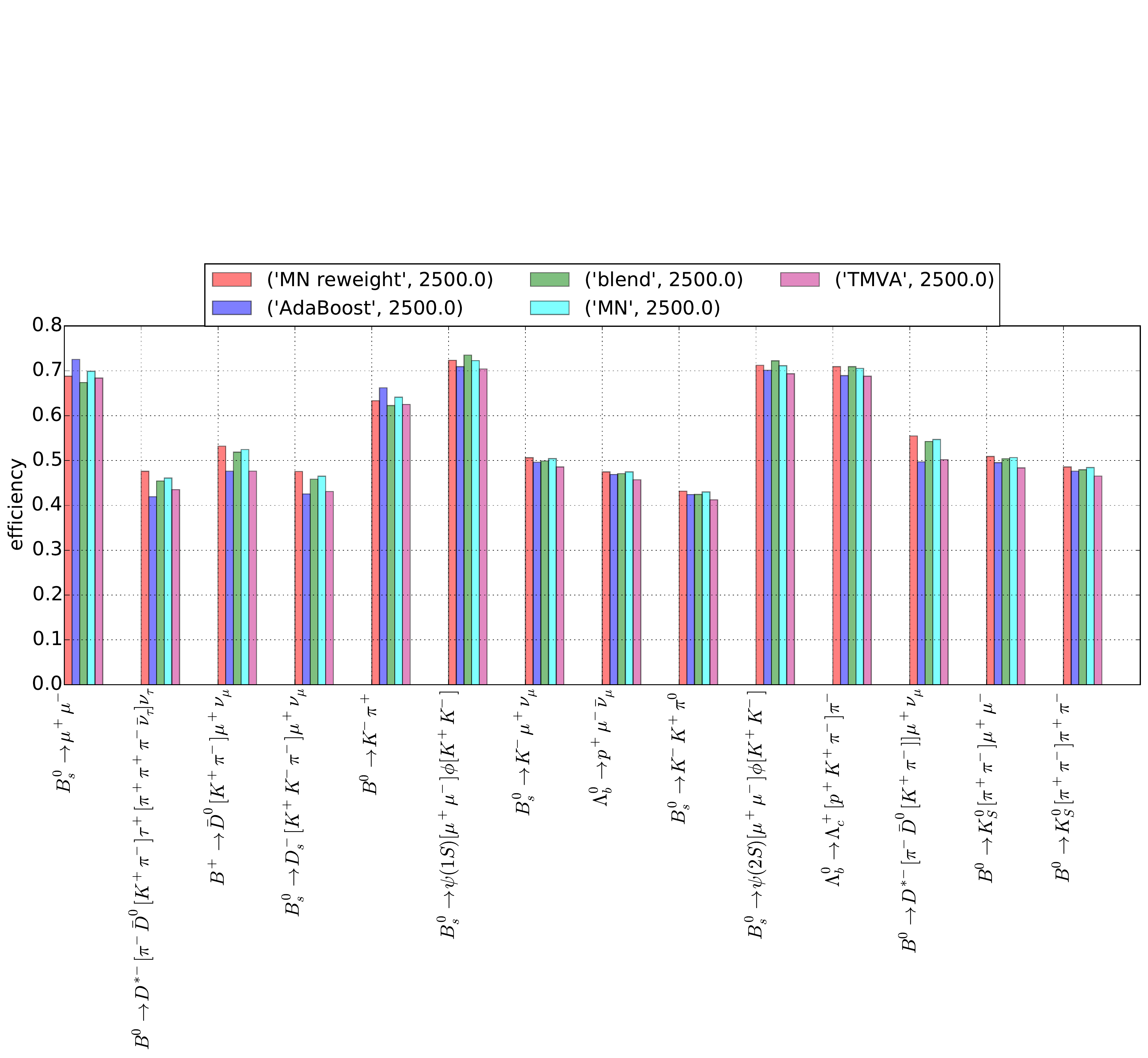}
\end{minipage} 
\caption{\label{hlt2_blend} Comparison of different algorithms: MatrixNet (MN), MatrixNet with SV weights (MN reweight), scikit-learn AdaBoost implementation (AdaBoost), TMVA AdaBoost implementation (TMVA) and some hierarchical algorithm (blend).}
\end{figure}

Different boosted decision trees were trained (figure~\ref{hlt2_blend}). Also some hierarchical algorithm was conducted, a so-called `blend'. Training data was divided into two parts. On the first part three MatrixNet classifiers were trained, which use only 2, 3 or 4 body decays as inputs. Next, the second part was predicted by the corresponding n-body classifiers. These predictions are considered as new additional input variables. The resulting MatrixNet was trained on this second part using basic variables and these additional ones. This hierarchical algorithm gives an improvement for several training modes.

Based on these studies, the MatrixNet (MN) model was chosen. Its efficiency was compared with the Run 1 algorithm. For these six training modes, we obtain significant relative improvement: 15-60\% for 2.5 kHz output rate and 50-80\% for 4 kHz (table~\ref{hlt2res}).

\begin{center}
\begin{table}[h]
\caption{\label{hlt2res} Ratio of Run-2 over Run-1 for HLT2/HLT1 efficiencies. 
Note that the denominator is reconstructible with $PT(B)>2$~GeV, $\tau(B)>0.2$~ps.}
%\footnotesize\rm
\centering
\begin{tabular}{@{}*{7}{l}{l}}
\br
mode & 2.5 kHz & 4. kHz \\
\mr
    $B^0\to K^*[K^+\pi^-]\mu^+\mu^-$ & 1.64 & 1.72   \\ 
    $B^+\to \pi^+K^-K^+$ & 1.59 & 1.65 \\ 
    $B^0_s\to D_s^-[K^+K^-\pi^-]\mu^+\nu_\mu$ & 1.14 & 1.47 \\ 
    $B^0_s\to \psi(1S)[\mu^+\mu^-]K^+K^-\pi^+\pi^-$ & 1.62 & 1.71 \\ 
    $B^0_s\to D_s^-[K^+K^-\pi^-]\pi^+$ & 1.46 & 1.52 \\ 
    $B^0\to D^+[K^-\pi^+\pi^+]D^-[K^+\pi^-\pi^-]$ & 1.40 & 1.86  \\
\br
\end{tabular}
\end{table}
\end{center}

\section{Online processing}
\begin{figure}[h]
\begin{minipage}{18pc}
\includegraphics[width=18pc]{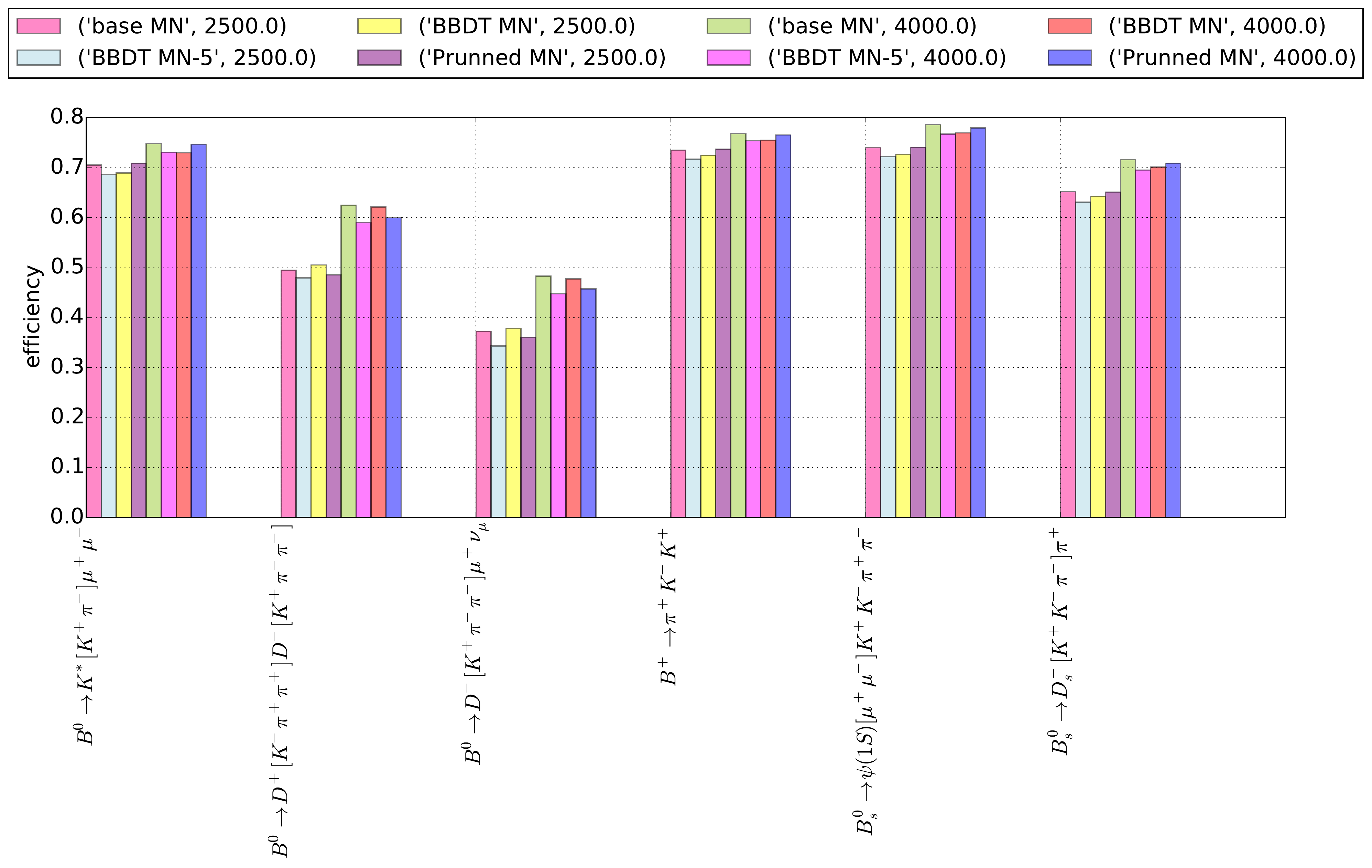}
\end{minipage}\hspace{2pc}%
\begin{minipage}{18pc}
\vspace{0.3cm}
\includegraphics[width=18pc]{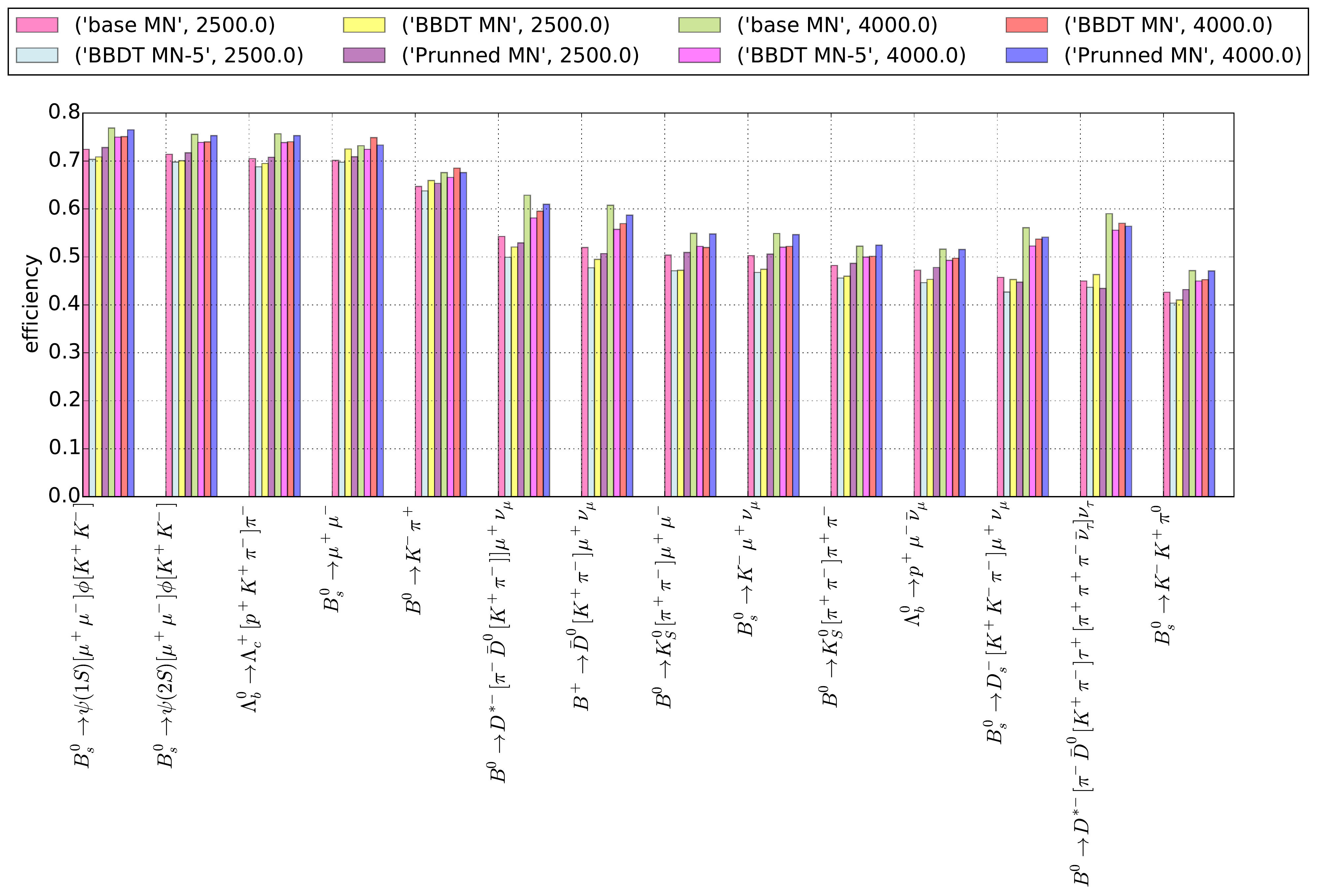}
\end{minipage} 
\caption{\label{hlt2_prun} Comparison of basic MatrixNet (base MN), BBDT format (BBDT MN) and post-prunned MatrixNet (Prunned MN)}
\end{figure}

Boosted decision trees are not appropriate for online processing events in triggers due to their low speed. Two approaches exist to overcome this restriction.   One is the so-called bonsai boosted decision format\cite{bbdt} (shortly BBDT).  Since it is a type of BDT, MatrixNet can be converted into this format. The second approach is post-prunning: the basic MatrixNet classifier includes several thousand trees, and the post-prunning procedure reduces this amount to a few hundred. This results in significant speedup of the prediction rate.  Both methods reduce MatrixNet signal efficiencies.  In case of the BBDT, we are also limited in the size of BBDT lookup table in RAM.  Different BBDT versions for MatrixNet were tried and compared to post-prunning to find the optimal solution for online processing (see figure~\ref{hlt2_prun}).

\begin{figure}[h]
\includegraphics[width=16pc]{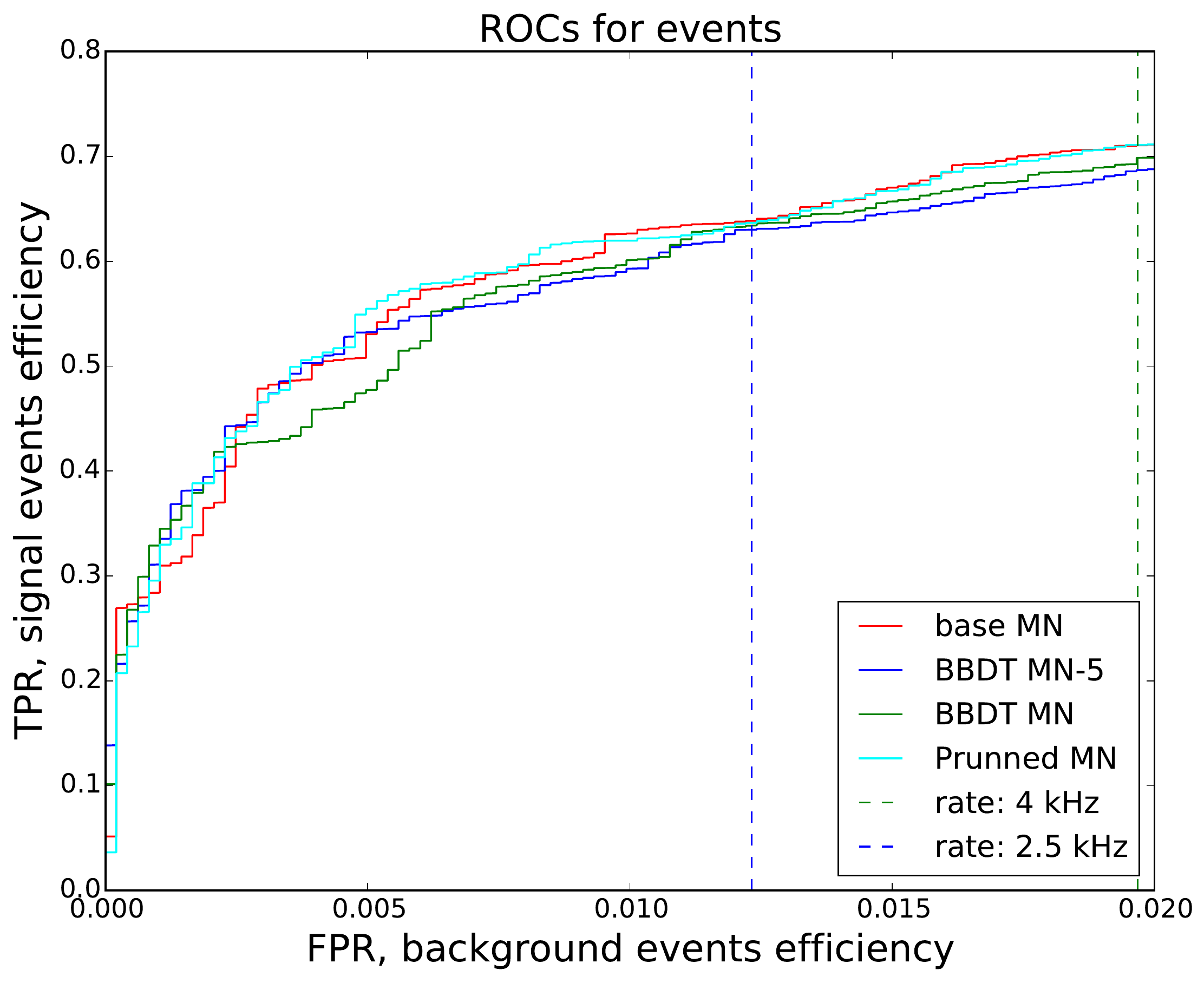}\hspace{2pc}%
\begin{minipage}[b]{22pc}\caption{\label{hlt2_prunroc} The best model among BBDT and post-prunning depends on background efficiency.}
\end{minipage}
\end{figure}

Interestingly, the preference between the BBDT and post-prunning depends on the chosen output rate. It is clear in figure~ \ref{hlt2_prunroc}, that the ROCs order depends on the background efficiency (or the output rate). Another study for Run 2 triggers is connected to a timing comparison of BBDT and post-prunning, which is in progress at this moment.

\section{Conclusion}
The LHCb topological trigger was successfully reoptimized for Run 2: 15-60\% efficiency improvement was obtained for 2.5 kHz output rate and 50-80\% for 4 kHz.  The presented HLT scheme will be applied in Run 2.

\end{document}